\documentclass[11pt]{article}
\usepackage{latexsym}
\usepackage[hypertex]{hyperref}
\usepackage{graphicx}

\def\lesssim{\mathrel{\mathpalette\vereq<}}
\def\gtrsim{\mathrel{\mathpalette\vereq>}}

\newcommand{\lsim}{\lesssim}

\makeatletter
\def\vereq#1#2{\lower3pt\vbox{\baselineskip1.5pt \lineskip1.5pt
\ialign{$\m@th#1\hfill##\hfil$\crcr#2\crcr\sim\crcr}}}
\makeatother

\begin{document}

\begin{flushright}
hep-ph/0402215 \\
\end{flushright}

\vspace{1.5cm}

\begin{center}
\mbox{\bf\LARGE Electroweak Symmetry Breaking} \\
\vspace*{4mm}
\mbox{\bf\LARGE via UV Insensitive Anomaly Mediation} \\

\vspace*{1.5cm}
{\large Ryuichiro Kitano, Graham D. Kribs, and Hitoshi
  Murayama\footnote{On leave of absence from Department of Physics,
    University of California, Berkeley, CA 94720.}} \\
\vspace*{0.5cm}
\mbox{\textit{School of Natural Sciences, Institute for Advanced Study, 
Princeton, NJ 08540}} \\

\vspace*{0.5cm}

\texttt{kitano@ias.edu, kribs@ias.edu, murayama@ias.edu}
\end{center}

\vspace*{1.0cm}

\begin{abstract}
  
Anomaly mediation solves the supersymmetric flavor and CP problems.
This is because the superconformal anomaly dictates that supersymmetry
breaking is transmitted through nearly flavor-blind infrared physics
that is highly predictive and UV insensitive.  Slepton mass squareds,
however, are predicted to be negative.  This can be solved by adding
$D$-terms for U(1)$_Y$ and U(1)$_{B-L}$ while retaining the
UV insensitivity.  In this paper we consider electroweak symmetry
breaking via UV insensitive anomaly mediation in several models.  For
the MSSM we find a stable vacuum when $\tan\beta < 1$, but in this
region the top Yukawa coupling blows up only slightly above the supersymmetry
breaking scale.  For the NMSSM, we find a stable electroweak breaking
vacuum but with a chargino that is too light. Replacing the cubic singlet term
in the NMSSM superpotential with a term linear in the singlet we find a
stable vacuum and viable spectrum.  Most of the parameter region with
correct vacua requires a large superpotential coupling, precisely what
is expected in the ``Fat Higgs'' model in which the superpotential is
generated dynamically.  We have therefore found the first viable
UV complete, UV insensitive supersymmetry breaking model that solves the
flavor and CP problems automatically: the Fat Higgs model with
UV insensitive anomaly mediation.  Moreover, the cosmological gravitino
problem is naturally solved, opening up the possibility of realistic 
thermal leptogenesis.

\end{abstract} 

\newpage
\section{Introduction}
\label{intro-sec}

The electroweak scale of the Standard Model is destabilized by
quadratically divergent radiative corrections to the Higgs mass squared.
It has long been known that supersymmetry provides an elegant resolution
to this naturalness problem by providing superpartners that cancel the
quadratic divergences to all orders in perturbation theory.  The absence
of experimental evidence for superpartners roughly requires that they
acquire supersymmetry breaking masses larger than of order the
electroweak scale.  The cancellation of the quadratic divergence is
therefore inexact, requiring a fine-tuning of about one part in $(m_{\rm
SUSY}/M_Z)^2$.  This fine-tuning is minimized when the size of soft
supersymmetry breaking is as close as possible to $M_Z$ \cite{finetune}.

The generic difficulty with such low superpartner masses is that they
can lead to large contributions to flavor changing neutral currents
(FCNC) and CP violating interactions at low energy.  Flavor-generic
supersymmetry breaking leads to effective FCNC and CP violating
operators that must be suppressed by tens to hundreds of TeV to avoid
conflict with experiment.  Raising the supersymmetry breaking scale
sufficiently does solve the supersymmetric flavor problem, but at the
expense of reintroducing a fine-tuning of at least one part in $10^6$.
Avoiding this fine-tuning problem requires carefully arranging
supersymmetry breaking to be mediated to the minimal supersymmetric
standard model (MSSM) in a flavor-blind way.\footnote{An alternative
possibility is to constrain the form of soft supersymmetry breaking
parameters using flavor symmetries.  However, the CP violation is
difficult to forbid unless the CP symmetry is exact at high energies and
is broken only spontaneously.}

There are only a few known flavor-blind mediation mechanisms: gauge
mediation \cite{gaugemediation}, anomaly mediation
\cite{Randall:1998uk,Giudice:1998xp}, and gaugino mediation
\cite{Kaplan:1999ac,Chacko:1999mi}.  Gauge mediation, gaugino mediation,
and several viable versions of anomaly mediation
\cite{Pomarol:1999ie,Chacko:1999am,Katz:1999uw,Bagger:1999rd,Jack:2000cd,Kaplan:2000jz,Chacko:2001jt,Okada:2002mv,Nelson:2002sa,Anoka:2003kn}
rely on specific UV physics to ensure flavor-blindness and an acceptable
spectrum.  Gauge mediation generically requires a messenger sector to
which the MSSM is coupled only through gauge fields.  Gaugino mediation
requires an extra dimension (or deconstruction
\cite{Csaki:2001em,Cheng:2001an}) with only gauge fields propagating in
the bulk.  Both frameworks require that the physics that explains the
origin of flavor, such as Froggatt--Nielsen mechanism
\cite{Froggatt:1978nt}, must be well above the mediation scale such that
the supersymmetric flavor problem is not reintroduced.  Anomaly
mediation requires a mechanism to suppress supergravity contributions
(sequestering across a bulk \cite{Randall:1998uk} or from a CFT
\cite{Luty:2001jh,Luty:2001zv}) as well as a mechanism to generate
positive slepton mass squareds.  A qualitatively different approach,
UV insensitive anomaly mediation \cite{Jack:2000cd,Arkani-Hamed:2000xj}
is the idea that soft breaking depends only on infrared (IR) physics
through anomaly mediation (sequestered using a bulk
\cite{Arkani-Hamed:2000xj} or CFT \cite{Harnik:2002et}) and $D$-terms
for hypercharge and $B-L$ (see also
\cite{Carena:2000ad,Murakami:2003pb}).

UV insensitive supersymmetry breaking is highly predictive since all the
supersymmetry breaking terms except for the $D$-terms can be calculated
with known low energy coupling constants.  The UV implementation of this
model requires sequestering, to prevent the usual supergravity
contributions to scalar masses, as well as a mechanism to generate
$D$-terms \cite{Jack:2000cd,Arkani-Hamed:2000xj,Harnik:2002et}.  The
virtue of this approach, however, is that the soft breaking masses are
independent of these details of the UV physics.  In this paper we
carefully examine electroweak symmetry breaking in the MSSM and
extensions with a singlet superfield with supersymmetry breaking
communicated via UV insensitive anomaly mediation.

This paper is organized as follows.  In Sec.~\ref{anom-sec} we review
anomaly mediation and the UV insensitive model with $D$-terms.  In
Sec.~\ref{MSSM-sec} we consider UV insensitive anomaly mediation with
the MSSM particle content.  We find that a minimum of the potential is
obtained only when the top Yukawa coupling is large, of order three, and
the ratio of Higgs vacuum expectation values ($\tan \beta = \langle H_2
\rangle / \langle H_1 \rangle $) is much less than 1.  The large top
Yukawa coupling runs into a Landau pole immediately above the supersymmetry
breaking scale.  The MSSM with UV insensitive anomaly mediation is
therefore UV incomplete and must be replaced with an unknown theory
above that scale.  This does not cause an inconsistency of our framework
because the soft terms are, by definition, UV insensitive.  However, the
large top Yukawa coupling prevents us from obtaining an accurate
estimate of the mass of the lightest Higgs boson due to the importance
of higher order radiative corrections.

We then consider extensions of the MSSM particle content by adding a
singlet superfield in the hopes of finding a UV complete and calculable
model.  In Sec.~\ref{NMSSM-sec} we first consider the next-to-minimal
supersymmetric standard model (NMSSM) with the superpotential $W=\lambda
S H_1 H_2 + (h/3) S^3$, where $S$ is the singlet superfield and $H_1$
and $H_2$ are the doublet Higgs superfields.  We find that this model is
not viable because the chargino mass is predicted to be much too small.
Then in Sec.~\ref{fat-sec} we consider an NMSSM-like model with
superpotential $W = \lambda S H_1 H_2 + m^2 S$ and $\lambda \gtrsim 1$.
The superpotential coupling $\lambda$ runs into a Landau pole at an
intermediate scale, implying that this model requires a UV completion
just like the MSSM.  Fortunately we know precisely what the UV
completion of this superpotential is: the ``Fat Higgs'' model
\cite{fat}!  The Fat Higgs model is a theory with SU(2) gauge group and
three (or four) flavors that gets strong at an intermediate scale.
These flavors confine into composite mesons with a dynamically generated
superpotential that is precisely of the form given above.  What is
perhaps most fascinating is that we find the parameter region with a
stable electroweak breaking vacuum occurs for values of the couplings
that are quite ordinary from the viewpoint of the Fat Higgs model.
Furthermore, the supersymmetric CP problem is solved automatically, and
the traditional $\mu$-problem is solved.  (A new $\mu$-like problem
resurfaces in the three-flavor Fat Higgs model, but unlike other
flavor-blind models, the solution to this new $\mu$-problem does
\emph{not} reintroduce the supersymmetric CP problem.)  Thus, the Fat
Higgs model is the first calculable UV complete model with
UV insensitive anomaly mediation that automatically solves both the
supersymmetric flavor and CP problems.

\section{Anomaly Mediation}
\label{anom-sec}

We first review supersymmetry breaking via anomaly mediation
\cite{Randall:1998uk,Giudice:1998xp}.  The supergravity Lagrangian can
be obtained from a local superconformal theory by a gauge fixing of
extra symmetries 
\cite{Cremmer:1978hn, Cremmer:1982en, Kugo:cu}.
The gauge fixing can be done by setting the values
of the components of a compensator chiral multiplet $\Phi$.  By
construction, $\Phi$ couples to the violation of conformal symmetry,
i.e., any dimensionful parameters such as mass parameters and also the
renormalization scale.  The scalar component of $\Phi$ is determined
such that the gravity kinetic term has the canonical normalization.
Supersymmetry breaking in the hidden sector causes the auxiliary
component $F_{\Phi}$ to acquire a non-vanishing value to cancel the
cosmological constant.  Since the vacuum energy from supersymmetry
breaking is its own order parameter, $F_{\Phi}$ is proportional to the
gravitino mass $m_{3/2}$.  With an appropriate redefinition of $\Phi$,
the scalar component can be fixed to be 1 and $F_{\Phi} = m_{3/2}$ in
this normalization.

Anomaly mediation is the contribution to the soft terms originated from
$F_{\Phi}$. The classical Lagrangian is given by
\begin{eqnarray}
 {\cal L} &=&
\int d^4 \theta
\left[
Q_i^\dagger e^{-2 V} Q_i
\right]
+ \left(
\int d^2 \theta
\frac{1}{2g^2} 
{\rm Tr} \left[
W^\alpha W_\alpha
\right]
+ h.c.
\right)
\nonumber \\
&&
-
\left(
\int d^2 \theta
\left[
\lambda_{ijk} Q_i Q_j Q_k
+ \Phi m_{ij} Q_i Q_j
+ \Phi^2 v^2_i Q_i
\right]
+ h.c.
\right) \ ,
\end{eqnarray}
where $Q_i$ and $V$ are the chiral and vector superfields, and
$W^\alpha$ is the field strength made of $V$.
We can read off the tree-level soft terms by substituting $\Phi = 1 + m_{3/2}
\theta^2$ into the above Lagrangian, resulting in
\begin{equation}
{\cal L}_{\rm soft} = \left(- B m_{ij} q_i q_j - C v^2_i q_i + h.c. \right)
\end{equation}
where $B=m_{3/2}$, $C=2 m_{3/2}$, and $q_i$ is the scalar component of
the chiral superfield $Q_i$.  Since the compensator also couples to the
renormalization scale, $\mu$, there are additional contributions at
quantum level.  The $\mu$ dependence appears in the wave function
renormalization and the gauge coupling as follows:
\begin{eqnarray}
 {\cal L} &=&
\int d^4 \theta
\left[ Z_i ( \frac{\mu}{\sqrt{\Phi \Phi^\dagger}})
Q_i^\dagger e^{-2 V} Q_i
\right]
+ \left(
\int d^2 \theta
\frac{1}{2g^2 (\frac{\mu}{\Phi})} 
{\rm Tr} \left[
W^\alpha W_\alpha
\right]
+ h.c.
\right)
\nonumber \\
&&
-
\left(
\int d^2 \theta
\left[
\lambda_{ijk} Q_i Q_j Q_k
+ \Phi m_{ij} Q_i Q_j
+ \Phi^2 v^2_i Q_i
\right]
+ h.c.
\right) \ .
\label{lagrangian-quantum}
\end{eqnarray}
Expanding $Z_i$ and $g^2$ in terms of $\theta^2$ and redefining the
fields to have canonical kinetic terms, we obtain trilinear scalar
couplings ($A_{ijk}$), scalar
mass squareds ($\tilde{m}^2$), and a gaugino mass $(m_\lambda)$ as
follows\footnote{ The gaugino mass parameter $m_\lambda$ flips its sign
when we redefine the gauginos $\lambda \to i \lambda$ in order to get
rid of the factor of $i$ in front of the gaugino-scalar-fermion vertex.
}:
\begin{eqnarray}
 A_{ijk} = - \lambda_{ijk} ( \gamma_i + \gamma_j + \gamma_k ) m_{3/2}\ ,\ \ \ 
 \tilde{m}_i^2 = \frac{1}{2} \dot{\gamma}_i m_{3/2}^2\ ,\ \ \ 
 m_\lambda = \frac{\beta}{g} m_{3/2}\ ,
\label{soft-terms}
\end{eqnarray}
where the anomalous dimension $\gamma_i$ and the beta function $\beta$ are
defined by
\begin{eqnarray}
 \gamma_i = - \frac{1}{2} \frac{d}{dt} \log Z_i (\mu)\ ,\ \ \ 
 \dot{\gamma}_i = \frac{d}{dt} \gamma_i\ ,\ \ \ 
 \beta = \frac{d}{dt} g\ , \ \ \ 
 t = \log \mu\ .
\end{eqnarray}
The soft parameters above are defined by the Lagrangian
\begin{eqnarray}
 {\cal L}_{\rm soft} =
- ( A_{ijk} q_i q_j q_k + h.c.)
- \tilde{m}^2_{i} |q_i|^2
- \frac{1}{2} m_{\lambda} \bar{\lambda} \lambda \ .
\end{eqnarray}
The results in Eq.~(\ref{soft-terms}) are true at any energy scale since
the form of the Lagrangian (\ref{lagrangian-quantum}) is always valid.
This indicates that the soft terms at a low-energy scale depend only on
the anomalous dimensions or beta functions at that scale and do not care
about the theory at higher energies.  This UV insensitivity is the main
feature of the anomaly mediation and implies that the soft terms are
calculable and nearly flavor-blind since the gauge contribution
overwhelms the Yukawa coupling contribution for all but the top squarks.

The problem is that slepton mass squareds are predicted to be negative.
This can be easily seen by an explicit calculation with the formula in
Eq.~(\ref{soft-terms}).  Neglecting the Yukawa interactions of the 
leptons, the slepton masses are given by
\begin{eqnarray}
 m_{\tilde{l}}^2 = \left(
- \frac{1}{2} b_Y g_Y^4 - \frac{3}{2} b_2 g_2^4
\right) M^2
\ ,\ \ \ 
 m_{\tilde{e}^c}^2 = 
- 2 \, b_Y g_Y^4 M^2 
\end{eqnarray}
where $M = m_{3/2} / (4 \pi)^2$.  The slepton masses depend overwhelmingly 
on the breaking of conformal symmetry by the gauge beta functions that
are given by
\begin{equation}
\beta_a = \frac{b_a g_a^3}{16 \pi^2} \; ,
\label{rge-gauge-eq}
\end{equation}
where the beta function coefficients [$b_Y$,$b_2$,$b_3$] are [$11$,$1$,$-3$]
for [U(1)$_Y$, SU(2)$_L$, SU(3)$_c$] gauge symmetries.
Upon inserting these numbers into the above expressions for the
slepton masses one immediately sees that the slepton mass squareds 
are negative, and therefore ``pure'' anomaly mediation is excluded.

For anomaly mediation to be viable there must be an additional
contribution to the slepton masses.  Additional contributions
generically involve UV sensitive physics
\cite{Pomarol:1999ie,Chacko:1999am,Katz:1999uw,Bagger:1999rd,Jack:2000cd,Kaplan:2000jz,Chacko:2001jt,Okada:2002mv,Nelson:2002sa,Anoka:2003kn},
that can lead to a perfectly viable supersymmetry breaking model.
However, the anomaly-mediated soft mass predictions, including UV
insensitivity, are generally lost.  An interesting alternative
possibility is to add $D$-term contributions to the sfermion masses
\cite{Jack:2000cd} that have been shown to preserve UV insensitivity
\cite{Arkani-Hamed:2000xj}.  If some non-MSSM fields charged under a
U(1) gauge symmetry acquire vacuum expectation values (VEVs) at a
slightly deviated point from the $D$-flat direction due to a
supersymmetry breaking effect, the $D$-term generates an additional
contribution to sfermion masses proportional to the U(1) charges through
the coupling of $-g {\cal Q}_i D |q_i|^2$ in the Lagrangian, where $g$
and ${\cal Q}_i$ are the gauge coupling constant and the charge of the
$q_i$ field, respectively.  The UV insensitivity was shown to be
preserved when the U(1) symmetry is anomaly free with respect to the
standard model gauge group.  In the MSSM there are two candidates of the
anomaly free U(1) symmetries, i.e., U(1)$_Y$ and U(1)$_{B-L}$, and those
$D$-term contributions are sufficient to resolve the tachyonic slepton
problem as we see below.  Although U(1)$_Y$ is unbroken above the
electroweak scale, the kinetic mixing between U(1)$_{B-L}$ and U(1)$_Y$
induces a $D$-term for U(1)$_Y$ \cite{Dienes:1996zr}, once a $D$-term
for U(1)$_{B-L}$ is generated.

The sfermion masses in UV insensitive anomaly mediation can therefore
be expressed as a ``pure'' anomaly mediated piece plus $D$-terms.
Evaluating the beta functions and gauge couplings at the weak 
scale,\footnote{We evaluated at the scale $\mu = 500$ GeV
for illustration, but threshold corrections should be taken into
account for accurate calculations of the physical masses.}
neglecting the Yukawa couplings, we find the first and second 
generation soft masses to be \cite{Arkani-Hamed:2000xj}
\begin{eqnarray}
  m^2_{\tilde{l}} &=& -0.344 M^2 + \frac{1}{2} D_{Y} + D_{B-L},
\nonumber \\
  m^2_{\tilde{e}^c} &=& -0.367 M^2 - D_{Y} - D_{B-L},
\nonumber \\
  m^2_{\tilde{q}} &=& 11.6 M^2 - \frac{1}{6} D_Y - \frac{1}{3} D_{B-L},
\\
  m^2_{\tilde{u}^c} &=& 11.7 M^2 + \frac{2}{3} D_Y + \frac{1}{3} D_{B-L},
\nonumber \\
  m^2_{\tilde{d}^c} &=& 11.8 M^2 - \frac{1}{3} D_Y + \frac{1}{3} D_{B-L},
\nonumber
\end{eqnarray}
while for the third generation scalars and Higgs bosons
\begin{eqnarray}
  m^2_{\tilde{l}_3} &=& -0.346 M^2 + \frac{1}{2} D_{Y} + D_{B-L},
\nonumber \\
  m^2_{\tilde{e}^c_3} &=& -0.371 M^2 - D_{Y} - D_{B-L},
\nonumber \\
  m^2_{\tilde{q}_3} &=& 9.40 M^2 - \frac{1}{6} D_Y - \frac{1}{3} D_{B-L},
\nonumber \\
  m^2_{\tilde{u}^c_3} &=& 7.37 M^2 + \frac{2}{3} D_Y + \frac{1}{3} D_{B-L},
\\
  m^2_{\tilde{d}^c_3} &=& 11.8 M^2 - \frac{1}{3} D_Y + \frac{1}{3} D_{B-L},
\nonumber \\
  m^2_{H_1}       &=& -0.395 M^2 + \frac{1}{2} D_Y,
\nonumber \\
  m^2_{H_2}       &=& -6.79  M^2 - \frac{1}{2} D_Y,
\nonumber
\end{eqnarray}
for $\tan\beta = 3$ as an example.

There is a window for the sleptons to be non-tachyonic:
\begin{equation}
  D_Y < - D_{B-L} < \frac{1}{2} D_Y < 0.
\label{D(B-L)range}
\end{equation}
Also, notice that the $D$-term contributions to the left-handed and the
right-handed up-type squarks are always negative, so that the $D$-terms
cannot be much larger than the anomaly-mediated contribution to ensure
all squark mass squareds stay positive.

For later use we evaluate the soft masses for $\tan \beta = 0.3$, 
\begin{eqnarray}
  m^2_{\tilde{l}} &=& -0.344 M^2 + \frac{1}{2} D_{Y} + D_{B-L},
\nonumber \\
  m^2_{\tilde{e}^c} &=& -0.367 M^2 - D_{Y} - D_{B-L},
\nonumber \\
  m^2_{\tilde{q}} &=& 11.6 M^2 - \frac{1}{6} D_Y - \frac{1}{3} D_{B-L},
\nonumber \\
  m^2_{\tilde{u}^c} &=& 11.7 M^2 + \frac{2}{3} D_Y + \frac{1}{3} D_{B-L},
\nonumber \\
  m^2_{\tilde{d}^c} &=& 11.8 M^2 - \frac{1}{3} D_Y + \frac{1}{3} D_{B-L},
\nonumber \\
  m^2_{\tilde{l}_3} &=& -0.344 M^2 + \frac{1}{2} D_{Y} + D_{B-L},
\label{sample-0.3}
 \\
  m^2_{\tilde{e}^c_3} &=& -0.368 M^2 - D_{Y} - D_{B-L},
\nonumber \\
  m^2_{\tilde{q}_3} &=& 550 M^2 - \frac{1}{6} D_Y - \frac{1}{3} D_{B-L},
\nonumber \\
  m^2_{\tilde{u}^c_3} &=& 1090 M^2 + \frac{2}{3} D_Y + \frac{1}{3} D_{B-L},
\nonumber \\
  m^2_{\tilde{d}^c_3} &=& 11.8 M^2 - \frac{1}{3} D_Y + \frac{1}{3} D_{B-L},
\nonumber \\
  m^2_{H_1}       &=& -0.342 M^2 + \frac{1}{2} D_Y,
\nonumber \\
  m^2_{H_2}       &=& 1678 M^2 - \frac{1}{2} D_Y.
\nonumber
\end{eqnarray}
The large top Yukawa coupling gives a huge enhancement in
$m^2_{\tilde{q}_3}$, $m^2_{\tilde{u}^c_3}$, and $m^2_{H_2}$.

\section{Electroweak Symmetry Breaking in the MSSM}
\label{MSSM-sec}

We now discuss electroweak symmetry breaking in the MSSM with 
UV insensitive ($D$-term modified) anomaly mediation.  We will find a 
parameter region with the correct electroweak breaking vacuum 
(the $Z$-boson mass is correct), but $\tan\beta < 1$.
The top Yukawa coupling is so large in the region that the coupling runs
into a Landau pole just above the supersymmetry breaking scale.

The stationary conditions of the potential can be solved analytically at
tree-level.  The conditions are well known
\begin{eqnarray}
 m_Z^2 &=& - \frac{ m_{H_1}^2 - m_{H_2}^2 }{\cos 2 \beta}
- ( m_{H_1}^2 + m_{H_2}^2 + 2 \mu^2 )
\ , \\
 \sin 2\beta &=& - \frac{2 B \mu}{m_{H_1}^2 + m_{H_2}^2 + 2 \mu^2 }
\ ,
\end{eqnarray}
where $\mu$ is the mass parameter in the superpotential $W = \mu H_1 H_2$. 
Since the $\mu$ parameter explicitly breaks conformal symmetry, 
the $B$ parameter is fixed to be $(4 \pi)^2 M = m_{3/2}$ as
we found in the previous section.  We solve the equations for $\mu$ and $D_Y$
as follows:
\begin{eqnarray}
 \mu &=& \frac{1}{2 \sin 2 \beta}
\left[
-B \pm \sqrt{
B^2 - 2 ( \bar{m}_{H_1}^2 + \bar{m}_{H_2}^2 ) \sin^2 2 \beta
}
\right]\ ,
\label{mu-formula} \\
 D_Y &=& 
\cos 2 \beta
\left[
- m_Z^2 - ( \bar{m}_{H_1}^2 + \bar{m}_{H_2}^2 + 2 \mu^2 )
\right]
- ( \bar{m}_{H_1}^2 - \bar{m}_{H_2}^2 )
\ ,
\label{DY-formula}
\end{eqnarray}
where $\bar{m}_{H_1}^2$ and $\bar{m}_{H_2}^2$ are the contributions from
pure anomaly mediation (without $D$-terms).
Since the $B$-term is enhanced by a factor of $(4 \pi)^2$ compared to
the soft masses, we expand the solution in terms of $1/(4 \pi)^2$.
Neglecting all the coupling constants except for the top Yukawa coupling and the
strong coupling constants, we obtain two solutions [corresponding to the
two signs of Eq.~(\ref{mu-formula})] that are approximately:
\begin{eqnarray}
\mbox{Solution 1:} & & \mu \sim - \frac{ \bar{m}_{H_2}^2 \sin 2 \beta}{ 2 B } 
\quad , \quad D_Y \sim \bar{m}_{H_2}^2 ( 1 - \cos 2 \beta ) 
\label{solution1-eq} \\
\mbox{Solution 2:} & & \mu \sim - \frac{B}{\sin 2 \beta} 
\quad , \qquad\quad D_Y \sim - \frac{ 2 B^2 \cos 2 \beta }{ \sin^2 2 \beta }
\label{solution2-eq}
\end{eqnarray}
Notice that the $\mu$ parameter is either ${\cal O}(M/(4 \pi)^2)$ or 
${\cal O}((4 \pi)^2 M)$ for Solution 1 or 2, respectively.

There are two further conditions for a stable electroweak breaking
vacuum.  The first condition is that $D_Y$ must be negative to stabilize 
the slepton direction [see Eq.~(\ref{D(B-L)range})], i.e., the sleptons 
must have positive mass squareds.
The second condition arises due to the lack of a quartic potential in 
the $D$-flat direction and therefore the quadratic terms must have 
positive coefficients,
\begin{eqnarray}
 \bar{m}_{H_1}^2 + \bar{m}_{H_2}^2 + 2 \mu^2 - 2 | B \mu | > 0
\ ,
\label{stability}
\end{eqnarray}
This ensures the potential is bounded from below.
For Solution 1 given in Eq.~(\ref{solution1-eq}), this condition 
reduces to
\begin{eqnarray}
 \bar{m}_{H_2}^2 
- |  \bar{m}_{H_2}^2  \sin 2 \beta | > 0\ ,
\end{eqnarray}
which is satisfied when $\bar{m}_{H_2}^2 > 0$.  Positive 
$\bar{m}_{H_2}^2$ can occur once the top Yukawa coupling becomes 
asymptotically non-free (requiring $\tan\beta \lesssim 1.5$).
But, positive $\bar{m}_{H_2}^2$ is in conflict with the first condition 
since the sign of $D_Y$ is positive for any $\tan\beta$.

Solution 2 always satisfies the stability condition in
Eq.~(\ref{stability}) and the sign of $D_Y$ is negative for $\tan\beta <
1$.  However, this solution has its own potential problem.  $D_Y$ is
generically very large, ${\cal O}((4 \pi)^4 M^2)$, as long as
$\tan\beta$ is not very close to 1, and this causes the squarks to get
negative mass squareds.  We numerically searched for a solution
corresponding to Solution 2 with $\tan\beta$ near 1 including the
one-loop corrections to the effective potential and we could not find a
solution.  Even if there was a solution, the lightest Higgs mass
vanishes at tree-level in the limit $\tan\beta \to 1$ and it would
be difficult to satisfy the experimental lower bound from the direct
searches of the Higgs boson.  Therefore, so long as the $1/(4 \pi)^2$
expansion is a good approximation, we find no acceptable solution.

Intriguingly, we did find a solution with $\tan\beta$ very small,
${\cal O}(0.3)$.  In this case the top Yukawa coupling $f_t = m_t / (v
\sin \beta)$ ($v=174$ GeV) is large and thus $f_t^4 / (4 \pi)^2$ is no
longer a good expansion parameter.  For large $f_t$, the
$\bar{m}_{H_2}^2$ parameter is given by
\begin{eqnarray}
 \bar{m}_{H_2}^2 \sim 18 f_t^4 M^2\ ,
\end{eqnarray}
and hence we cannot neglect the $2 \bar{m}_{H_2}^2 \sin^2 2 \beta$
factor in Eq.~(\ref{mu-formula}).  Similarly, one can show that $D_Y$
does become negative for $\tan\beta \lesssim 0.3$.
Adding the one-loop corrections to the effective potential is not
qualitatively different from the tree-level analysis.  An example
solution is
\begin{eqnarray}
 M = 300 {\rm ~GeV}\ ,\ \ 
 \mu = -4251  {\rm ~GeV}\ ,\ \ 
 \tan \beta = 0.29\ ,\ \
 D_Y = - 2.5 M^2\ ,\ \ 
 D_{B-L} = 2.0 M^2\ ,
\label{sample_MSSM}
\end{eqnarray}
that has a proper electroweak symmetry breaking vacuum.

The small value for $\tan\beta = 0.29$ causes the top Yukawa coupling to
be large, $f_t(M_Z) = 3.5$.  Such a large top Yukawa coupling is
quasi-perturbative and runs into a Landau pole just above the
supersymmetry breaking scale (mass of the top squarks).  The
perturbative MSSM description breaks down and must be replaced by new
description such as a theory with composite top quarks and/or Higgs
fields.  In the low-energy effective theory, the theory above the cutoff
scale is of course unknown.  Nevertheless, the UV insensitivity of
anomaly mediation allows us to calculate the soft breaking parameters in
the IR, and therefore does not lose any predictability at low energies.

It is interesting to estimate the lightest Higgs boson mass ($m_h$) for
the set of parameters in Eq.~(\ref{sample_MSSM}) to clarify whether
$m_h$ is greater than the present experimental lower bound of 114.4 GeV
\cite{EWWG}.  We calculated $m_h$ with the one-loop correction from the
(s)top loop diagrams.  Although $m_h$ is small at tree-level, the
one-loop correction is huge because the top squarks are very heavy,
8.6~TeV and 12~TeV, due to the large top Yukawa coupling.
The resulting lightest Higgs mass turns out to be 176 GeV.\footnote{
  We ignored the momenta of the external lines in the calculation.}
However, the uncertainly of the calculation is large because of the
heavy stops and large $f_t$.  Although our one-loop analysis is unable
to give a definite prediction, we suspect it is not inconsistent with
the experimental lower bound.  Nevertheless a more extensive (beyond
one-loop) analysis is necessary to be definitive.

\section{Electroweak Symmetry breaking in the NMSSM?}
\label{NMSSM-sec}


We have found an electroweak symmetry breaking solution in the MSSM, but
the model is not UV complete due to the large top Yukawa coupling
blowing up just about the supersymmetry breaking scale.  The difficulty
in the MSSM is caused by the large value of the $B$ parameter that is
induced through the explicit breaking of the conformal symmetry by the
$\mu$ parameter.  This suggests we should consider a model with no
explicit mass parameters, such as the NMSSM \cite{Ellis:1988er} with
superpotential
\begin{eqnarray}
 W = \lambda S H_1 H_2 + \frac{h}{3} S^3\ .
\label{W-NMSSM}
\end{eqnarray}
Since all the soft breaking parameters are loop suppressed, we
naturally expect that no large values of the coupling constants
will be needed.  The $\mu$ parameter in the MSSM is replaced by 
$\lambda \langle S \rangle$ that generates the Higgsino mass.
Despite these good features, we were unable to find a viable set of
parameters because the Higgsino mass turns out to be too small.

The origin of the difficulty in the NMSSM is again due to the high 
predictability of anomaly mediation.  The scalar mass parameter for 
the singlet $m_S^2$ is always positive.  This is because $S$ interacts 
only through the couplings $h$ and $\lambda$, where $h$ is always
asymptotically non-free and $\lambda$ is asymptotically non-free 
unless it is much smaller than the gauge couplings. 
Therefore $S \sim 0$ is almost a stable point of the potential.
In order for $S$ to acquire a sizable VEV, large values of either 
the linear or the cubic terms of $S$ in the scalar potential are needed. 
Such terms can be obtained from trilinear scalar couplings $A_{\lambda}$
and $A_{h}$, but in anomaly mediation these terms will be large only if
$\lambda$ and $h$ are also large.  Large superpotential couplings 
$\lambda$ and $h$ induce a larger $m_S^2$, that then suppresses the VEV.  
We therefore find that an arbitrarily large VEV of $S$ cannot be obtained 
due to this correlation induced by anomaly mediation.

\begin{figure}[t]
\hspace*{3.5cm}
\includegraphics[width=7.5cm]{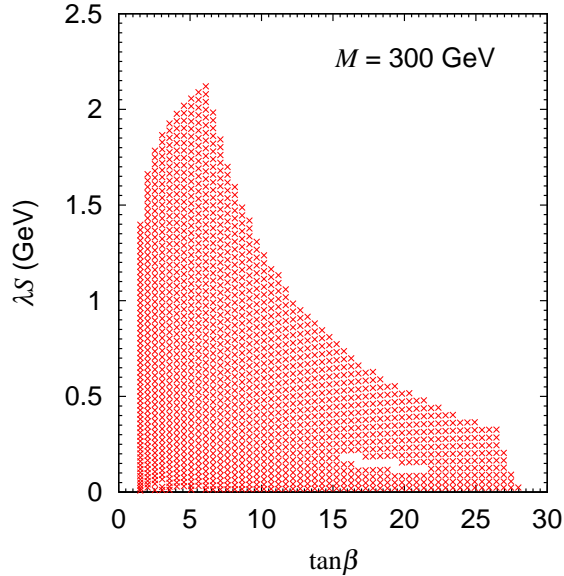} 
\caption{The region of parameters where a stable electroweak symmetry 
breaking vacuum is found in the NMSSM.  The Higgsino mass parameter 
$\mu = \lambda S$ is too small $\protect\lesssim 3$ GeV at all the points.
}
\label{fig:tanbeta-mu}
\end{figure}

We numerically searched for a parameter set by solving stationary
conditions.  In Fig.~\ref{fig:tanbeta-mu} we show solutions found with 
$M=300$~GeV in the ($\tan\beta$, $\lambda S$)-plane.  Upon fixing $M$,
$\tan\beta$, and $\lambda S$ there are no remaining free parameters in
the Higgs sector.  In particular, all the solutions shown in the Figure
have a small $\lambda S \lesssim 2.2$ GeV leading to a tiny chargino 
mass that is excluded by experimental bounds.
Increasing $M$ is of no help 
since the soft mass $m_S^2 \propto M^2$ increases faster than 
the trilinear scalar coupling $A_\lambda \propto M$.

One possible way to save the NMSSM is to couple the singlet to
additional vector-like matter in the superpotential $\lambda' S \Phi
\overline{\Phi}$ \cite{Agashe:1997kn,deGouvea:1997cx}.  The new
superpotential coupling $\lambda'$ could be asymptotically free if the
contribution from the strong coupling dominates.  It would be
interesting to verify whether this works or not, but the addition of
vector-like matter is somewhat {\it ad hoc}\/ and so we will not pursue
this further in this paper.

\section{Electroweak Symmetry Breaking in the Fat Higgs model}
\label{fat-sec}

The difficulty of generating a large VEV for $S$ in the NMSSM suggests
that a different form of the superpotential is needed.  In particular,
consider instead
\begin{equation}
  W = \lambda S H_1 H_2 + m^2 S\ ,
  \label{W-fat_Higgs}
\end{equation}
where $S$ is again a singlet under the SM gauge interactions.  
This superpotential is the same as the NMSSM superpotential (\ref{W-NMSSM}) 
with the replacement $k S^3/3 \to m^2 S$.  The linear term forces $S$ 
to acquire a large VEV through the soft breaking linear term 
$C m^2 S$ (``$C$-term'') in the scalar potential.

Interestingly, this superpotential is precisely the low energy effective
superpotential of the Minimal Supersymmetric Fat Higgs model \cite{fat}.
The Fat Higgs model is a very recent idea in which electroweak symmetry
is dynamically broken as a result of a new gauge interaction SU(2)$_H$
getting strong at an intermediate scale $\Lambda_H$.  The minimal field
content in the UV theory is three flavors (six doublets) $T^{1 \ldots
6}$ that is known to confine and generate a dynamical superpotential
\cite{Seiberg:1994bz}
\begin{eqnarray}
 W_{\rm dyn.} = \frac{{\rm Pf} M}{\Lambda_H^3} \ ,
\end{eqnarray}
for the low energy effective theory of the meson composite fields 
$M_{ij} \sim T^i T^j$.  The separation between the confinement scale
and the electroweak breaking scale is controlled by adding a
supersymmetric mass term for one flavor ($m_T T^5 T^6$).  Other
superpotential terms are added ensuring that the vacuum aligns to the
one with proper electroweak breaking and additional spectator meson composites
are lifted.  The resulting low energy effective 
superpotential is then
\begin{eqnarray}
 W_{\rm eff.} = \lambda M_{56} (M_{14} M_{23} - M_{24} M_{13} - v_0^2)\ ,
\end{eqnarray}
where $v_0 \simeq \sqrt{m_T \Lambda_H}/4 \pi$.  
This superpotential is identical 
to Eq.~(\ref{W-fat_Higgs}) upon identifying $M_{56} \to S$, 
$(M_{14}, M_{24}) \to H_1$, $(M_{13}, M_{23}) \to H_2$ and 
$- \lambda v_0^2 \sim m^2$.  
The superpotential (\ref{W-fat_Higgs}) is therefore UV complete even if 
the superpotential couplings are large.  This will be crucial to be able
to find a viable parameter region with electroweak symmetry breaking 
and no large fine-tuning.

\subsection{The Scalar Potential}
\label{potential-subsec}

We first discuss the scalar potential and solve the constraints explicitly.
In the supersymmetric limit, the superpotential (\ref{W-fat_Higgs})
forces the VEVs
\begin{equation}
 \langle H_1^0 \rangle = \langle H_2^0 \rangle = \sqrt{\frac{- m^2}{\lambda}}, 
\qquad
  \langle S \rangle = 0.
\end{equation}
In the presence of the soft supersymmetry breaking effects, the full
potential is given by (setting the charged components to zero)
\begin{eqnarray}
  \label{eq:V}
  V &=& | \lambda H_1^0 H_2^0 + m^2|^2
  + | \lambda |^2 |S|^2 (|H_1^0|^2 + |H_2^0|^2) 
  + m_{H_1}^2 |H_1^0|^2
  + m_{H_2}^2 |H_2^0|^2 
  + m_S^2 |S|^2 \nonumber \\
  & &{} + \left[ A_\lambda S H_1^0 H_2^0 + C m^2 S + h.c.\right]
  + \frac{g_Y^2 + g_2^2 }{8} (|H_1^0|^2 - |H_2^0|^2)^2.
\end{eqnarray}
In general $\lambda$ and $m^2$ are complex parameters, but we can absorb
the phases of these parameters by rotations of $S$ and
$H_1$ (or $H_2$).\footnote{This induces a shift in the phase of the determinant of
the Yukawa matrices that amounts to shifting the strong CP phase.  The
strong CP problem is beyond the scope of this work.}  Similarly, phase
of the soft breaking mass $m_{3/2}$ can be absorbed by a U(1)$_R$
rotation.  Hence, there are no remaining physical phases that are not
proportional to the CKM phase or absorbed into the definition of the
strong phase, and therefore this model does not have a supersymmetric CP
problem.  It is possible that the expectation values of the Higgs fields
could have a non-vanishing phase that causes spontaneous CP violation,
but we assume we are living in a (sufficiently long-lived
\cite{Kusenko:1996jn}) local vacuum that does not break CP.  We
therefore restrict our discussion to fields with real VEVs of $H_1^0$,
$H_2^0$, and $S$.

The soft terms are expressed in terms of the anomalous dimensions by the
formulae in Eq.~(\ref{soft-terms}).  In the Fat Higgs model, the anomalous 
dimension factors are given by
\begin{eqnarray}
  \gamma_{H_1} &=& \frac{1}{(4\pi)^2} (\lambda^2 -
  \frac{3}{2} g_2^2 - \frac{1}{2} g_Y^{2}), \\
  \gamma_{H_2} &=& \frac{1}{(4\pi)^2} (\lambda^2 + 3f_t^2 -
  \frac{3}{2} g_2^2 - \frac{1}{2} g_Y^{2}), \\
  \gamma_S &=& \frac{1}{(4\pi)^2} 2\lambda^2, \\
  \gamma_{q_3} &=& \frac{1}{(4\pi)^2} (f_t^2 - \frac{8}{3}g_3^2 -
  \frac{3}{2}g_2^2 - \frac{1}{18} g_Y^{ 2}), \\
  \gamma_{u^c_3} &=& \frac{1}{(4\pi)^2} (2f_t^2 - \frac{8}{3}g_3^2 -
  \frac{8}{9}g_Y^{2}),
\end{eqnarray}
where $g_Y$, $g_2$, and $g_3$ are the gauge coupling constants of
U(1)$_Y$, SU(2)$_L$, and SU(3)$_C$, respectively, while $q_3$
and $u_3^c$ represent the left-handed and right-handed up-type quarks of
the third generation.
The one-loop renormalization group equations for the couplings are
given by,
\begin{eqnarray}
  \frac{d}{dt} \lambda &=& \lambda (\gamma_{H_1} + \gamma_{H_2} +
  \gamma_S), \\
  \frac{d}{dt} f_t &=& f_t (\gamma_{H_2} + \gamma_{q_3} +
  \gamma_{u_3^c}),
\end{eqnarray}
and by Eq.~(\ref{rge-gauge-eq}) just as in the MSSM.
Inserting these beta functions into the expressions for the soft masses,
we obtain
\begin{eqnarray}
  m_{H_1}^2 &=& \left( 4 \lambda^4 + 3\lambda^2 f_t^2 - \lambda^2
    (3g_2^2 + g_Y^{2}) - \frac{3}{2} g_2^4 - \frac{11}{2} g_Y^{4}
  \right) M^2
  + \frac{D_Y}{2},\\
  m_{H_2}^2 &=& \left( 4\lambda^4 + 6\lambda^2 f_t^2 + 18f_t^4 - \lambda^2
    (3g_2^2 + g_Y^{2}) - f_t^2 (16 g_3^2 + 9 g_2^2 + \frac{13}{3}
    g_Y^{2}) \right. \nonumber \\
  & & \left. - \frac{3}{2} g_2^4 - \frac{11}{2} g_Y^{4}
  \right) M^2
  - \frac{D_Y}{2}, \label{mH2-formula} \\
  m_S^2 &=& \left(8\lambda^4 + 6\lambda^2 f_t^2 - \lambda^2 (6g_2^2 +
    2g_Y^{2})\right) M^2, \\
  A_\lambda &=& - \lambda \left(4\lambda^2 + 3f_t^2 - 3g_2^2 - g_Y^{2}\right) M, \\
  C &=& 2 (4\pi)^2 M.
\end{eqnarray}
The $C$-term is $(4 \pi)^2$ enhanced because the linear term for $S$ in
the superpotential violates the conformal symmetry at tree-level.

There are three minimization conditions to determine the VEVs:
\begin{eqnarray}
 A_\lambda S + \lambda m^2 
&=& - \frac{ \sin 2 \beta }{2}
(m_1^2 + m_2^2 + \lambda^2 v^2)\ ,
\label{constraint-1} \\
 m_Z^2 &=& - \frac{m_1^2 - m_2^2}{\cos 2 \beta} - ( m_1^2 + m_2^2 )\ ,
\label{constraint-2} \\
0 &=& S ( \lambda^2 v^2 + m_S^2 ) + \frac{1}{2} A_\lambda v^2 \sin 2 \beta + 
C m^2 ,
\label{constraint-3}
\end{eqnarray}
where $m_1^2 \equiv m_{H_1}^2 + \lambda^2 S^2$, $m_2^2 \equiv
m_{H_2}^2 + \lambda^2 S^2$, and $v = \sqrt{ \langle H_1^0 \rangle^2 +
\langle H_2^0 \rangle^2 }$.  Setting $v=174$~GeV, we can solve for 
$S$, $D_Y$, and $m^2$ as a function of $\tan\beta$.  We obtain
\begin{eqnarray}
  \lefteqn{
    S = \frac{1}{4 \lambda C}
    \Bigg[ 
      \left( m_S^2 + 
      \lambda^2 v^2 - \frac{A_\lambda C}{\lambda} \right) 
    \frac{2}{\sin 2 \beta} }
  \nonumber \\
  & &{} \pm \sqrt{ 
      \left( m_S^2 + \lambda^2 v^2 - \frac{A_\lambda C}{\lambda} \right)^2 
\frac{4}{\sin^2 2 \beta}
      + 8 A_\lambda C \lambda v^2  
      - 8 C^2 ( \bar{m}_{H_1}^2 + \bar{m}_{H_2}^2 + \lambda^2 v^2 ))}\Bigg]
 \ ,
\label{S-fat_Higgs}
\end{eqnarray}
where $\bar{m}_{H_1}^2$ and $\bar{m}_{H_2}^2$ are the pure anomaly-mediated
soft masses (without $D$-terms).  
For each solution for $S$, we determine $D_Y$ and $m^2$ as
\begin{eqnarray}
 D_Y &=& - \cos 2 \beta \left[
m_Z^2 + \bar{m}_{H_1}^2 + \bar{m}_{H_2}^2 + 2 \lambda^2 S^2
\right] - ( \bar{m}_{H_1}^2 - \bar{m}_{H_2}^2 )\ ,
\label{DY-fat_Higgs} \\
 m^2 &=& - \frac{\sin 2 \beta}{2 \lambda}
( \bar{m}_{H_1}^2 + \bar{m}_{H_2}^2 + 2 \lambda^2 S^2 + \lambda^2 v^2  )
- \frac{A_\lambda S}{\lambda}\ .
\label{m2-fat_Higgs}
\end{eqnarray}
$D_Y$ must be negative to give a positive contributions to both of the 
slepton soft mass squareds.  For $\tan\beta > 1$, we find the first term 
in Eq.~(\ref{DY-fat_Higgs}) is positive for $\lambda \sim {\cal O}(1)$ 
while the 
second term can be sufficiently negative because of the contribution from 
the strong coupling in Eq.~(\ref{mH2-formula}).  
For $\tan \beta < 1$, the first term is negative while the second term 
can be positive.

\begin{figure}[t]
\includegraphics[width=7.5cm]{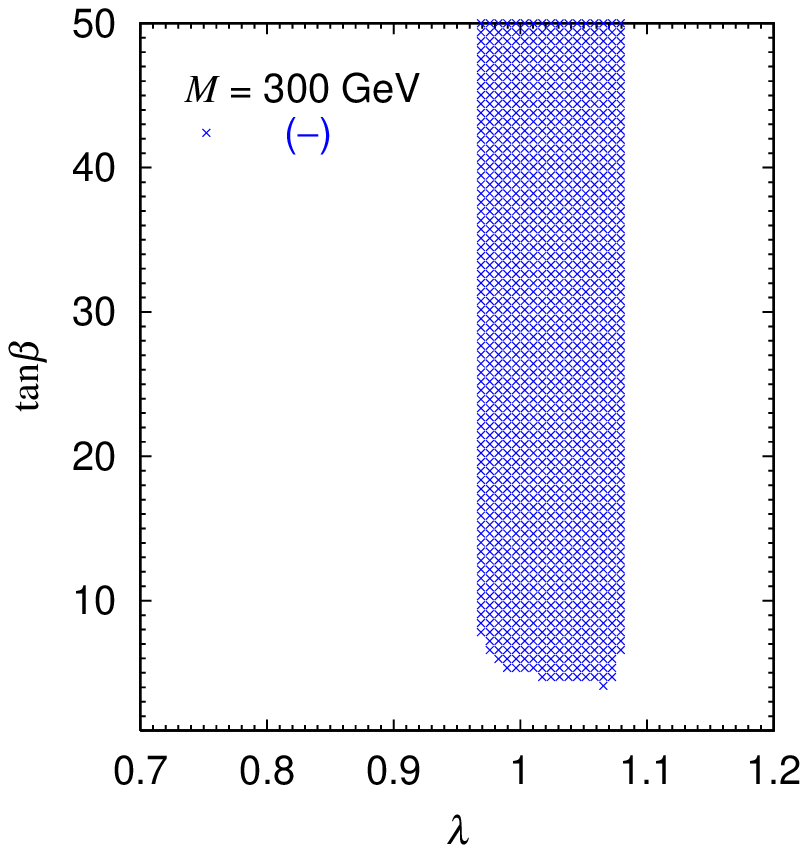} 
\includegraphics[width=7.5cm]{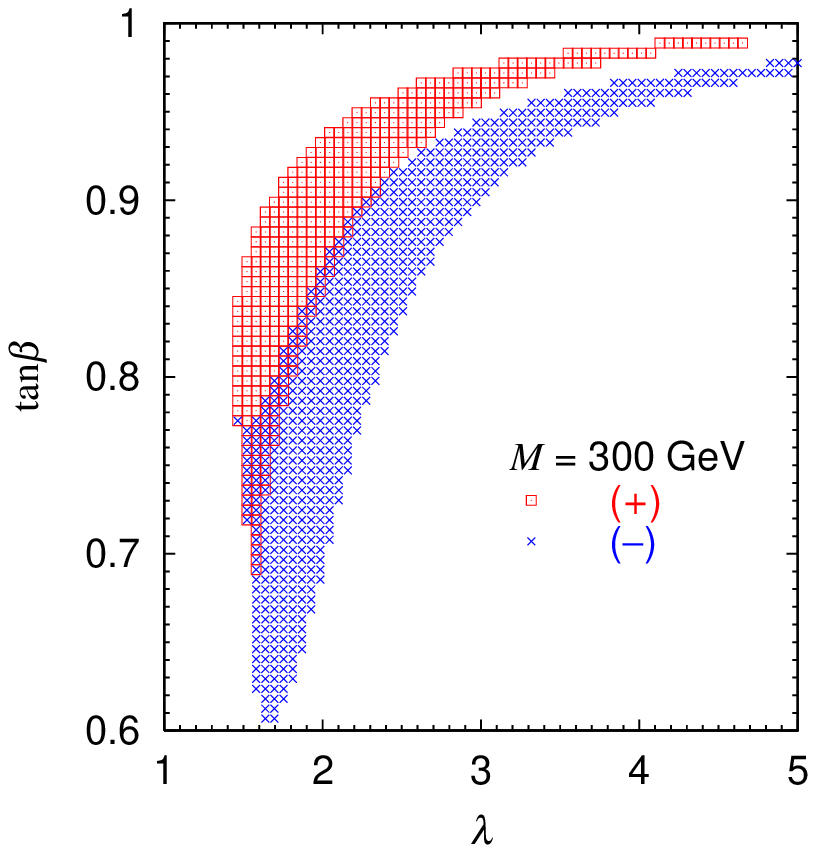} 
\caption{
The parameter regions in the Fat Higgs model where electroweak symmetry
breaking solutions with all scalar particle mass squareds positive is shown
for fixed $M = 300$~GeV. 
The two plots correspond to the two distinct regions of parameters
$\tan\beta > 1$ and $\tan\beta < 1$.  For $\tan\beta < 1$, there are 
two solutions shown by $\Box$ and $\times$ that correspond to the
positive and negative sign solution to Eq.~(\ref{S-fat_Higgs}),
respectively.
} 
\label{fig:lam-tanbeta}
\end{figure}

We have four parameters in the Higgs sector ($\lambda$, $m^2$, $M$,
$D_Y$). One combination of these parameters is fixed by the $Z$ mass.
This leaves three independent parameters that we choose to be ($M$,
$\lambda$, $\tan \beta$) with the remainder of the parameters determined
by Eqs.~(\ref{S-fat_Higgs})--(\ref{m2-fat_Higgs}).  In
Fig.~\ref{fig:lam-tanbeta}, the parameter region where solutions exist
in the $(\lambda, \tan \beta)$-plane are shown for a fixed value of $M$.
All scalar field mass squareds were required to be positive by choosing
the $D_{B-L}$ parameter appropriately.  As anticipated above, there are
two separate regions with $\tan \beta > 1$ and $\tan \beta < 1$.
Similar regions are obtained for other values of $M$.  We should
emphasize that it is quite non-trivial to obtain sets of parameters with
a stable electroweak breaking vacuum, especially given that there are
only three parameters in the model.


The $\tan\beta > 1$ case corresponds to the negative-sign solution of 
Eq.~(\ref{S-fat_Higgs}) while the positive-sign solutions are excluded 
because they imply $D_Y > 0$.  Fig.~\ref{fig:lam-tanbeta} (left plot) 
clearly shows that $\lambda$ is fixed to a narrow range around 1.  The
lower and upper limits on $\lambda$ can be understood as follows.  A
small $\lambda$ implies that the pseudo-scalar Higgs mass parameter
$m_1^2 + m_2^2 + \lambda^2 v^2$ is small, and this causes a (physical)
Higgs mass squared to go negative as we will see in
Sec.~\ref{mass-subsec}.
The upper bound on $\lambda$ is due to the requirement $D_Y < 0$.  Also,
negative $D_Y$ requires $f_t$ to be sufficiently asymptotically free.
This gives a lower limit of about $\tan \beta \gtrsim 5$.

For $\tan\beta <1$, solutions are found for both signs of
Eq.~(\ref{S-fat_Higgs}).  Fig.~\ref{fig:lam-tanbeta} (right plot) 
shows that $1.4 \lesssim \lambda \lesssim 5$.  This region of $\lambda$ 
is precisely in the range that corresponds to the Fat Higgs model with
compositeness at an intermediate scale.  The scale of compositeness can
be estimated by running $\lambda$ up to its Landau pole.
We find it fascinating that
a typical value of $\lambda \sim 2$ is perfectly consistent with
supersymmetry breaking communicated via UV insensitive anomaly mediation 
in the Fat Higgs model.  As was stressed in \cite{fat}, there is nothing
wrong with a larger $\lambda$ or $f_t$ coupling that run into non-perturbative
values since the composite Higgs description is expected to break down 
at an intermediate scale.

\begin{figure}[t]
\includegraphics[width=15cm]{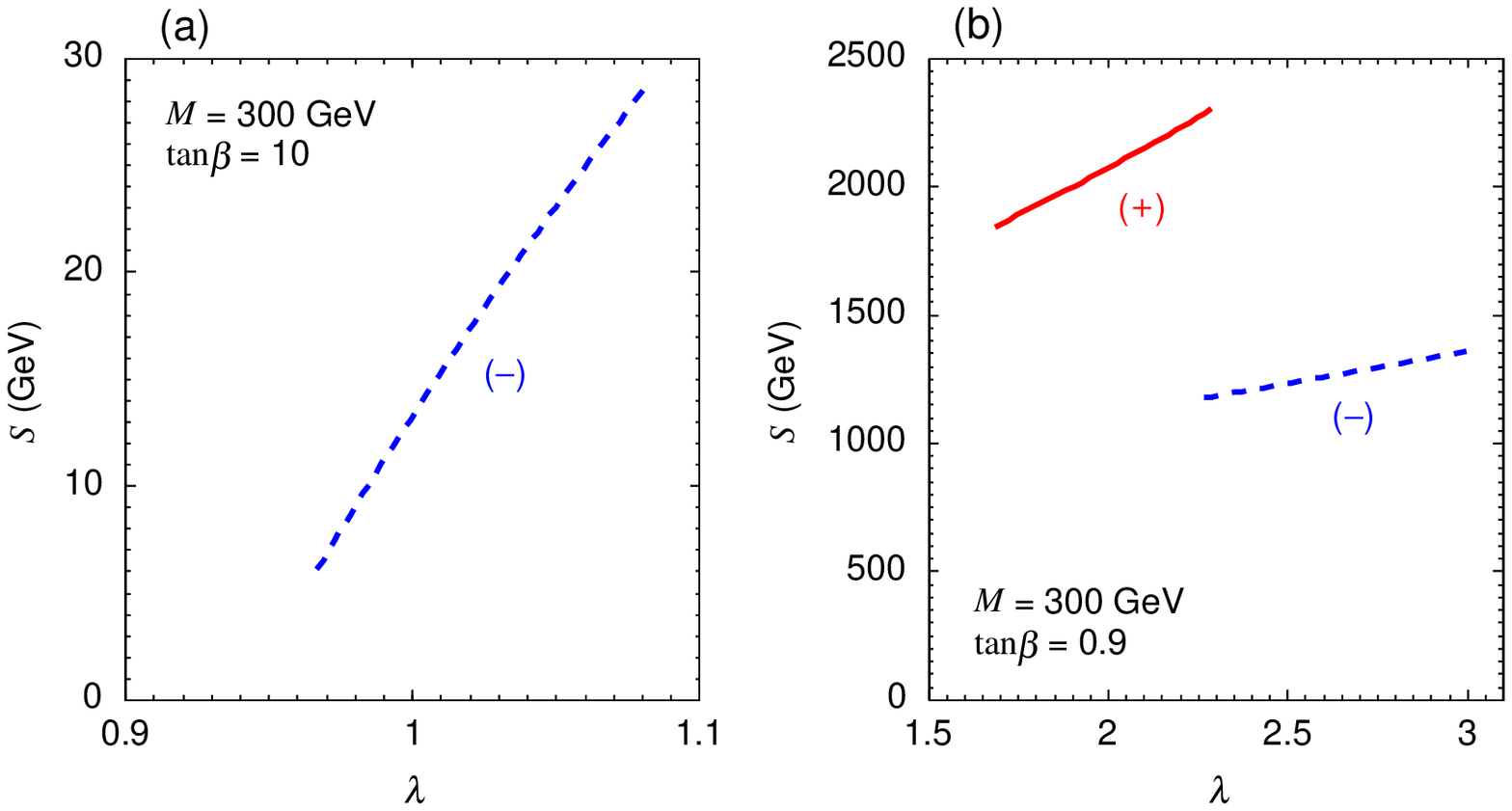}
\caption{
The expectation values of $S$ are shown as functions of $\lambda$
with fixed $M = 300$~GeV and $\tan\beta = 10$ [plot (a)] and 0.9 
[plot (b)].
} 
\label{fig:lam-dep}
\end{figure}

The corresponding values of $S$ for these solutions is shown in
Fig.~\ref{fig:lam-dep}.
For $\tan \beta < 1$, the VEV of $S$ is ${\cal O}$(TeV), and thus there 
is no light Higgsino problem (unlike the NMSSM).
The solution with $\tan \beta > 1$, however, does generically have a
light Higgsino problem.  From Eq.~(\ref{constraint-1}) it is clear that
small $S$ is a consistent solution because of the suppression of 
$\sin 2\beta$ when $\tan\beta$ is large.  The large values of $S$ are 
obtained only when $M$ is increased, but this also increases the 
fine-tuning of the Higgs potential for this solution.

\subsection{The Mass Spectrum}
\label{mass-subsec}

The Fat Higgs model with UV insensitive anomaly-mediated soft breaking 
has a mass spectrum that is distinctly different from the MSSM.
Here we present the mass matrices of the superpartners and discuss 
the general features of the spectrum.

We first consider the Higgs boson mass.  There are three CP-even Higgs
bosons in the model. The mass matrix in the basis ($H_1^0$,
$H_2^0$, $S$) is given by
\begin{eqnarray}
 \left(
\begin{array}{ccc}
 m_Z^2 \cos^2 \beta + m_X^2 \sin^2 \beta 
& *
& *\\
 - (m_X^2 + m_Z^2 - 2 \lambda^2 v^2) \sin \beta \cos \beta
& m_Z^2 \sin^2 \beta + m_X^2 \cos^2 \beta 
& * \\
( A_\lambda \sin \beta + 2 \lambda^2 S \cos \beta ) v
& ( A_\lambda \cos \beta + 2 \lambda^2 S \sin \beta ) v
& m_S^2 + \lambda^2 v^2 \\
\end{array}
\right)\ ,
\end{eqnarray}
where $m_X^2 = m_1^2 + m_2^2 + \lambda^2 v^2$ and the off-diagonal
entries $*$ are suppressed using the obvious symmetry of the matrix.  In
the limit where $M \gg v$ and/or $\lambda \gg 1$, we can ignore the
mixing between the singlet and the doublet Higgses.
%
%
If we consider the $2 \times 2$ sub-matrix of the doublet
Higgses, the eigenvalues are given by
\begin{eqnarray}
  \lefteqn{
    m^2_{H^0, h^0} \sim
    \frac{1}{2}
    \bigg[
    m_Z^2 + m_X^2 }
    \nonumber \\
    & &{}
    \pm
    \sqrt{
      (m_Z^2 - m_X^2)^2
      + 4 m_Z^2 m_X^2 \sin^2 2 \beta
      - 4 \lambda^2 v^2 
      \left(
        m_Z^2 + m_X^2 + \lambda^2 v^2
      \right) \sin^2 2 \beta
    }
    \bigg]\ .
\end{eqnarray}
From the above formula, we see that the lightest Higgs mass approaches
$m_Z$ when $\tan\beta \gg 1$ or $\tan\beta \ll 1$.  Whether such a
parameter choice is phenomenologically viable depends on adding the
usual radiative corrections to the Higgs mass
\cite{Okada:1990vk,Ellis:1990nz,Haber:1990aw}.  For $\tan \beta \sim
1$, $m_{h^0}$ can be much larger than $m_Z$.  This is an interesting
feature of the Fat Higgs model.  

In the same basis, the mass matrix for the CP-odd Higgs bosons is given by
\begin{eqnarray}
 \left(
\begin{array}{ccc}
 m_X^2 \sin^2 \beta 
& m_X^2 \sin \beta \cos \beta & -A_\lambda v \sin \beta \\
 m_X^2 \sin \beta \cos \beta
& m_X^2 \cos^2 \beta & -A_\lambda v \cos \beta \\
 -A_\lambda v \sin \beta
&-A_\lambda v \cos \beta 
& m_S^2 + \lambda^2 v^2 \\
\end{array}
\right)\ .
\end{eqnarray}
After rotating the (12)-plane by the angle $\beta$, the matrix becomes
\begin{eqnarray}
 \left(
\begin{array}{ccc}
 0 & 0 & 0 \\
 0 & m_X^2 & - A_\lambda v \\
 0 & -A_\lambda v & m_S^2 + \lambda^2 v^2 \\
\end{array}
\right)\ .
\end{eqnarray}
The massless mode corresponds to the would-be Nambu-Goldstone boson
eaten by the $Z$ boson.
We can clearly see that the mass parameter $m_X^2$ has to be at least
positive in order for the potential to be stable. That gives the lower
limit of $\lambda$ for solutions with $\tan \beta > 1$.

The charged Higgs boson mass is given by
\begin{eqnarray}
 m_{H^\pm}^2 = m_1^2 + m_2^2 + m_W^2\ ,
\end{eqnarray}
that is reminiscent of the MSSM $m_{H^\pm}^2|_{\rm MSSM} = 2 \mu^2 +
m_{H_1}^2 + m_{H_2}^2 + m_W^2$.

Charginos have a similar mass matrix to that in the MSSM:
\begin{eqnarray}
 \left(
\begin{array}{cc}
 M_2 & - g_2 v \cos\beta \\
 - g_2 v \sin\beta & \lambda S \\
\end{array}
\right)\ ,
\end{eqnarray}
where $M_2$ is the gaugino mass of SU(2)$_L$ given by
$M_2 = - g_2^2 M$.
The charginos are heavier than the experimental bounds from LEP 
so long as the Higgsino mass parameter is $\lambda S \gtrsim
{\cal O}(100$~GeV), and the supersymmetry breaking parameter 
$M \gtrsim 250$~GeV.

The neutralino sector is very different from the MSSM.
There is an additional neutralino ($\psi_S$: singlino) from the fermionic
component of $S$.
The mass matrix is given in the basis ($\tilde{B}^0$, $\tilde{W}^0$,
$\tilde{H}_1^0$, $\tilde{H}_2^0$, $\psi_S$) as follows:
\begin{eqnarray}
 \left(
\begin{array}{ccccc}
 M_1 & *  & * & * & * \\
 0 & M_2 & * & * & * \\
 g_Y v \cos \beta / \sqrt{2} & - g_2 v \cos \beta / \sqrt{2} & 0 & * & * \\
 - g_Y v \sin \beta / \sqrt{2} & g_2 v \sin \beta / \sqrt{2} & - \lambda S & 0 & * \\
 0 & 0 & -\lambda v \sin \beta & - \lambda v \cos \beta & 0 \\
\end{array}
\right)\ ,
\end{eqnarray}
where $M_1= -11 g_Y^2 M$ is the gaugino mass of U(1)$_Y$.  The
singlino mass is induced only through the mixing with the Higgsinos.

The formulae for the sfermion masses are the same as the MSSM by
replacing the $\mu$ parameter with $\lambda S$.  Positive slepton
mass squareds require (\ref{D(B-L)range}) to be satisfied, and there 
is an upper bound on $|D_Y|$ to ensure the squark mass squareds do
not become negative.
An interesting feature is that the stops become much heavier than the
other sfermions when $\tan \beta < 1$, since the asymptotically non-free
top Yukawa coupling gives large positive contributions to the stop masses.
This is a distinctly different feature from other supersymmetry breaking 
scenarios.

\subsection{Examples of Viable Parameters}
\label{example-subsec}

Here we demonstrate explicitly the viability of UV insensitive anomaly
mediation with the Fat Higgs model superpotential, by presenting sample
sets of parameters with our computations of the soft breaking
parameters.  From Sec.~\ref{potential-subsec}, we know stable
electroweak breaking vacua exist for two separate regions of the
parameter space: $\tan\beta < 1$ and $\tan\beta > 1$.

%
%

\begin{table}
\begin{center}
\begin{tabular}{|lcl|l|l|} \hline\hline
          & \phantom{0000} &          & Point I            & Point II \\ 
          & &                         & ($\tan\beta < 1$)  & ($\tan\beta > 1$) 
\\ \hline
input     & & $\lambda$               & $2.3$              & $1.05$ \\
parameters & & $m^2$ [GeV$^2$]     & $-3.1 \times 10^5$ & $-4.8 \times 10^3$ \\
         & & $M$ [GeV]               & $300$              & 900 \\
         & & $D_Y$ [GeV$^2$]         & $-2.8 M^2$           & $-2.9 M^2$ \\
         & & $D_{B-L}$ [GeV$^2$]     & $2 M^2$            & $2 M^2$ \\ \hline
output   & & $\tan\beta$             & $0.9$              & $5.0$ \\
results  & & $S$ [GeV]               & 1186               & 133 \\
         & & $m_{H_1^0}$ [GeV]       & 355                & 47 \\
         & & $m_{H_2^0}$ [GeV]       & 4976               & 1930 \\
         & & $m_{H_3^0}$ [GeV]       & 6614               & 3258 \\
         & & $m_{A_1^0}$ [GeV]       & 4924               & 1926 \\
         & & $m_{A_2^0}$ [GeV]       & 6662               & 3260 \\
         & & $m_{H^\pm}$ [GeV]       & 6613               & 1954 \\
         & & $m_{\chi_1^\pm}$ [GeV]  & 128                & 140 \\
         & & $m_{\chi_2^\pm}$ [GeV]  & 2730               & 394 \\
         & & $m_{\chi_1^0}$ [GeV]    & 57                 & 44.6 \\
         & & $m_{\chi_2^0}$ [GeV]    & 128                & 214 \\
         & & $m_{\chi_3^0}$ [GeV]    & 419                & 241 \\
         & & $m_{\chi_4^0}$ [GeV]    & 2731               & 394 \\
         & & $m_{\chi_5^0}$ [GeV]    & 2785               & 1257 \\
         & & $m_{\tilde{l}_L}$ [GeV] & 147                & 423 \\
         & & $m_{\tilde{e}_R}$ [GeV] & 202                & 640 \\
         & & $m_{\tilde{\tau}_1}$ [GeV] & 144                & 422 \\
         & & $m_{\tilde{\tau}_2}$ [GeV] & 205                & 638 \\
         & & $m_{\tilde{u}_L}$ [GeV] & 1241               & 3724 \\
         & & $m_{\tilde{u}_R}$ [GeV] & 1209               & 3622 \\
         & & $m_{\tilde{t}_1}$ [GeV] & 1689               & 3026 \\
         & & $m_{\tilde{t}_2}$ [GeV] & 2131               & 3444 \\
         & & $m_{\tilde{d}_L}$ [GeV] & 1241               & 3725 \\
         & & $m_{\tilde{d}_R}$ [GeV] & 1313               & 3941 \\
         & & $m_{\tilde{b}_1}$ [GeV] & 1313               & 3441 \\
         & & $m_{\tilde{b}_2}$ [GeV] & 1728               & 3929 \\ 
\hline\hline
\end{tabular}
\end{center}
\caption{Two sample sets of parameters and the resulting superpartner
masses are shown.  Note that these are tree-level values; radiative 
corrections are important especially for the lightest Higgs mass.} 
\label{sample-table}
\end{table}

In Table~\ref{sample-table} we present sample parameters and the
computed spectra for $\tan\beta < 1$ (Point I) and $\tan\beta > 1$
(Point II).  The input parameters include the supersymmetric parameters
$\lambda$ and $m^2$, and the supersymmetry breaking parameters $M$,
$D_Y$ and $D_{B-L}$.  Everything is calculated from these five
parameters.  $\tan\beta$ and $S$ are fixed at the minimum of the
potential, while the remainder of Table~\ref{sample-table} provides the
masses of the superpartners.

Point I is perhaps the most interesting case.  At tree-level, the
lightest Higgs boson mass is much larger than $m_Z$ because of the
quartic term in the potential has a coefficient of $\lambda^2$
\cite{fat}.  The lightest supersymmetric particle (LSP) is the lightest
neutralino that is mostly a singlino, and thus the lower bound from the
direct search experiments does not apply.  Other than the lightest Higgs
boson, the particles in the Higgs sector are several TeV due to the
large value of $\lambda$.

The large $\lambda$ and $f_t$ indicate the existence of strong coupling
well above the electroweak scale.  This is precisely what is expected in
the Fat Higgs model when the Higgs and singlet superfields' compositeness
is revealed.  Indeed, large $\lambda$ and the existence of a linear
term in $S$ is a prediction of the low energy effective theory of 
the Fat Higgs model.

The second column shown as Point II in Table~\ref{sample-table} provides
a sample spectrum with $\tan\beta > 1$.  As was anticipated earlier, the
solution with $\tan\beta > 1$ requires large $M$ to have sufficiently
heavy Higgsinos.  Note that the tree-level mass for the lightest Higgs
boson of 47 GeV is rather light but is expected to be modified by a
large radiative correction from the (s)top-loop and $S$-loop diagrams.
Parameterizing the loop contributions as
\begin{eqnarray}
 m_{H^0_1}^2 = (m_{H^0_1}^{\rm tree} )^2 + \Delta_t + \Delta_S\ , 
\end{eqnarray}
we obtain
\begin{eqnarray}
 \Delta_t \sim \frac{3}{4 \pi^2} \left( \frac{m_t^4}{v^2} \right) \log
 \frac{m_{\tilde{t}}^2}{m_t^2} \sim (105{\rm ~GeV})^2 \ ,
\end{eqnarray}
that should be large enough to push the Higgs mass above the
LEP bound.

\subsection{Phenomenology}
\label{pheno-subsec}

%
%

The main general predictions of the anomaly mediation are: (1) A
relatively large difference in mass between the colored superpartners
(gluino, squarks) and the uncolored ones (sleptons, neutralinos,
charginos), (2) the characteristic prediction on the gaugino masses in
proportion to their beta functions, and (3) the heavy gravitino.
Next, we know the superpotential given in Eq.~(\ref{W-fat_Higgs}) is
UV completed by the Fat Higgs model \cite{fat} that also determines
electroweak symmetry breaking.  In this framework we found two
distinct regions of parameters represented by Points I and II given in
Table~\ref{sample-table} that we will use for the discussion of the
phenomenology.

For the region of parameters characterized by $\tan\beta \lsim 1$,
represented by Point I, we found (1) a large effective $\mu$-parameter,
and (2) a light neutralino that is mostly a singlino.  For the region of
parameters characterized by $\tan\beta \sim 5$, represented by Point II,
we found (1) the effective $\mu$-parameter at the weak scale, and (2) a
light neutralino that is mostly a Higgsino.  These general features lead
to interesting and quite distinct phenomenology and cosmology from
conventional supersymmetric models with minimal supergravity or its
slight modifications.

The first observation is that the Fat Higgs model with UV insensitive
anomaly mediation contains heavy squark and gluino masses yet
only a mild fine-tuning comparable to that in the MSSM 
(discussed in the next section).  Unlike the MSSM, however,
the flavor problem as well as the supersymmetric CP problem 
are completely solved due to the UV insensitivity.


The LHC is expected to discover supersymmetry for Point I through
missing energy and dilepton signatures of squarks and gluinos,
although their heaviness limits the event rate and hence precision
measurements will not be easy.  Nonetheless $M_3$ should be measured
fairly well \cite{Hinchliffe:1996iu}.  The chargino and the second
neutralino are basically winos, and the sleptons are light.  The
trilepton signature is a target at Tevatron.  At LHC, both the
trilepton signature of chargino/neutrlino and dilepton signature of
sleptons can be searched for.  For Point II, squarks and gluinos may
be beyond the reach of the LHC.  On the other hand, both chargino
states and four neutralino states are relatively light and can be
looked for.  In anomaly-mediated models studied in the literature with
the addition of a UV sensitive universal scalar mass, the lighter
chargino and the lightest neutralino are mostly wino and have very
small mass splitting .  In our UV insensitive model, by contrast, the
lighter chargino decays into the lightest neutralino with a sizable
mass splitting and thus the chargino signal becomes a good one to
seek.  

The uncolored states likely can be produced at an $e^+ e^-$ Linear
Collider of $\sqrt{s} = 0.5$--1~TeV.  The most important measurement
is the chargino and neutralino compositions and the extraction of
$M_1$ and $M_2$ \cite{Tsukamoto:1993gt}.  Combining the LHC and LC
data, it seems feasible to experimentally verify the main quantitative
prediction of anomaly mediation that is the ratios of gaugino masses
(it may leave an ambiguity with a particular type of gauge-mediated
supersymmetry breaking \cite{Kribs:1999df}, but this can be resolved
by studying the scalar masses).

The Higgs spectrum is quite distinct from conventional supersymmetry.
There is a Standard Model-like Higgs that is relatively heavy in Point
I.  In this mass range, the Higgs mostly decays into $WW$ and $ZZ$
that makes its discovery and further study quite easy at the LHC.  It
has another potential invisible decay mode into a pair of the lightest
neutralinos.  Fortunately, the lightest neutralino has little Higgsino
component and hence the invisible branching fraction is suppressed.
Therefore, the LHC can find not only the Higgs but also determine its
width, its spin, and couplings to $WW$ and $ZZ$ at 10\% level.  For
Point II, the tree-level Higgs mass is light but the radiative
correction pushes it above the current LEP bound.  In this case the
invisible decay mode into a pair of the lightest neutralino becomes
important, and the search may be difficult.  Nonetheless the
$W$-fusion process with double forward jet tagging will be helpful
\cite{Cavalli:2002vs}.  On the other hand, the Higgs would
definitely be found at an $e^+ e^-$ Linear Collider.  For both Points,
verifying the model experimentally is difficult because the other
Higgs states are all very heavy, in the multi-TeV range, and they are
only pair-produced.  The only potential sensitivity is in the
precision electroweak measurement, {\it i.e.}\/ at GigaZ, where an
additional contribution to the $T$-parameter from the heavy Higgs
states may be extracted.  Much higher energy colliders such as VLHC
and/or CLIC will be needed to directly probe these states.

The scalar mass spectrum is determined by only three parameters,
the overall scale and two $D$-terms, and thus has many non-trivial 
sum rules.  Gaugino masses determine $M$, while left- and right-handed 
sleptons determine two $D$-terms, and together with $m_H$, there are 
no remaining free parameters.  The remainder of the superpartner spectrum 
could then be predicted and searched for at colliders.
Therefore precise measurements of squark masses would allow
non-trivial tests of the model.  This is a task for a very high-energy
$e^+ e^-$ linear collider such as CLIC.

One potential concern is the invisible width of $Z$-boson for the Point
II, as $Z$-boson may decay into a neutralino pair.  For this particular
set of parameters the invisible width is modified by less than 1\% of an
additional neutrino species, and is completely acceptable.
Nevertheless, in general there is a significant constraint on the region
of parameters with $\tan\beta \gtrsim 5$ because we find the mass of the
lightest neutralino tends to be less than but of order $m_Z/2$.

Finally, a few comments on neutrinos and lepton flavor violation are
in order.  UV insensitive anomaly mediation naturally incorporates
right-handed neutrinos.  This is simply because three generations of
right-handed neutrinos are required to gauge U(1)$_{B-L}$ that itself
is needed to allow the $D$-term for U(1)$_{B-L}$.  There are several
possibilities for generating neutrino masses
\cite{Arkani-Hamed:2000xj,Harnik:2002et}:
\begin{itemize}
\item Forbid the usual $\lambda_\nu L H_2 \nu^c$ term in the
  superpotential but include the higher dimension operator
  $\lambda_\nu L H_2 \nu^c/M_{\rm Pl}$ in the K\"ahler potential
  \cite{Arkani-Hamed:2000xj}.  Inserting appropriate powers of the
  compensator and expanding this out one obtains a Yukawa coupling of
  order $m_{3/2}/M_{\rm Pl}$, allowing for Dirac neutrinos of about
  the right scale while preserving UV insensitivity.
\item The usual see-saw mechanism \cite{seesaw} with order one Yukawa
  couplings $\lambda_\nu L H_2 \nu^c$ and a large right-handed
  neutrino mass scale [U(1)$_{B-L}$ breaking scale] generated
  dynamically \cite{Harnik:2002et}.  This will induce UV sensitive
  corrections to the lepton sector of the low energy theory, but the
  leading order result remains unaffected.
\end{itemize}
If the usual see-saw mechanism is adopted, the deflection from the 
UV insensitive trajectory would lead to potentially observable lepton 
flavor violating processes, such as $\mu \rightarrow e\gamma$.
This is not required, of course, since the right-handed neutrino
mass scale could be much lower with smaller neutrino Yukawa couplings.

\subsection{Cosmology}
\label{cosmology-sec}

Several cosmological implications of anomaly mediation with UV
insensitivity are quite interesting.  The gravitino problem of
supersymmetry are practically solved because it decays before the Big
Bang Nucleosynthesis \cite{Gherghetta:1999sw,Moroi:1999zb}.  The
gravitino is expected to decay into the lightest supersymmetric particle
and other particles with hadronic or electromagnetic energy that
thermalize quickly.  Therefore, the only constraint is that the
gravitino is not too abundant before their decay so as not to overclose
the universe through the LSP decay product.  This consideration
($\Omega_\chi \leq 1$), applied to gravitinos, leads to an upper bound
on the reheating temperature \cite{Kawasaki:1994af}
\begin{equation}
  T_R \leq 2.7 \times 10^{11} {\rm GeV}
  \left(\frac{100~{\rm GeV}}{m_\chi}\right)
  h^2 .
\end{equation}
Rescaling to the dark matter density measured by WMAP, 
$\Omega h^2 \simeq 0.11$ \cite{Spergel:2003cb}
the revised constraint is
\begin{equation}
  T_R \leq 3.0 \times 10^{10} {\rm GeV}
  \left(\frac{100~{\rm GeV}}{m_\chi}\right) .
\end{equation}
Note that if the reheating temperature is close to the upper bound,
the neutralino can give the correct abundance even if the annihilation
cross section is too large.

Interestingly, the large allowed reheating temperature (due to the large
gravitino mass) allows thermal leptogenesis.  For a sub-TeV gravitino,
the reheating temperature is required to be less than $10^6$--$10^9$~GeV
\cite{Kawasaki:1994af}, while thermal leptogenesis prefers $T_{RH}
\gtrsim 10^{10}$~GeV \cite{Giudice:2003jh, Buchmuller:2004nz}.  Even
though strict UV insensitivity requires right-handed sneutrinos to have
a mass at the electroweak scale, they can be made heavy as in the
conventional seesaw mechanism with potentially observable lepton-flavor
violation as discussed in the previous subsection.  In this case, the
thermal leptogenesis would be viable.

As for dark matter, UV insensitive anomaly mediation predicts the LSP is
the lightest neutralino that is mostly a singlino or a Higgsino.  In
general this allows the mass of the LSP to be smaller in comparison with
conventional models.  The lightest neutralino acquires its mass due to a
see-saw like structure in the neutralino mass matrix.  To a good
approximation, we can ignore the mixing with the gaugino states for both
regions of parameters.  For Point I, the two neutral Higgsino states
have a vector-like mass of $\mu = \lambda S$, while the determinant in
the lower three-by-three matrix is $2\lambda^2 \mu v^2
\sin\beta\cos\beta$.  Therefore the lightest neutralino has a mass of
approximately $m_{\chi_1^0} \simeq 2\lambda^2 v^2
\sin\beta\cos\beta/\mu$ and its wave function is
\begin{equation}
  \chi_1^0 \simeq \tilde{S}
  -\frac{\lambda v \cos\beta}{\mu} \tilde{H}_u^0
  -\frac{\lambda v \sin\beta}{\mu} \tilde{H}_d^0.
\end{equation}
Despite the large $\mu$, however, the mixing is not so small given a
large $\lambda v$.  When $\tan\beta \approx 1$, the $H_u^0$ and $H_d^0$
components in $\tilde{\chi}_1^0$ cancel in the coupling to the
$Z$-boson, while the annihilation due to the Higgs exchange is also
small due to the large Higgs mass and its $P$-wave nature.  The
annihilation cross section is of the order of $10^{-12}$~GeV$^{-2}$,
while the correct abundance is obtained for $\langle\sigma_{\it ann}
v_{\it rel}\rangle \simeq10^{-9} {\rm GeV}^2$.  The neutralino is
therefore overabundant in Point I.  However, there are viable parameter
solutions in which the mass of the lightest neutralino is much closer to
$m_Z/2$, and therefore the annihilation cross section is significantly
enhanced by the $Z$-pole in the $P$-wave.  A completely separate
possibility is that a would-be overabundant singlino is diluted by
entropy production below the electroweak scale.  Conversely, for Point
II the lightest neutralino has a large Higgsino component (66.5\%) and a
singlino component (33.3\%).  Then the $Z$-exchange in the $s$-channel
is large due to the large number of final states despite its $P$-wave
nature.  The annihilation cross section is of the order of
$10^{-7}$~GeV$^{-2}$.  Obviously the precise relic abundance of any
particular set of parameters must be calculated carefully, but we are
happy to find that the abundance varies around the cosmologically
interesting value and the lightest neutralino ought to be a good dark
matter candidate.

The moduli fields expected in most hidden sector models of supersymmetry
breaking as well as string theory acquire their masses from
supersymmetry breaking of the order of $m_{3/2}$, and decay with a
lifetime comparable to that of the gravitino.  They are typically
displaced from the minimum with an ${\cal O}(M_{Pl})$ amplitude, 
dominating the universe before they decay, making Big Bang Nucleosynthesis 
(BBN) impossible.  They also
dilute any preexisting baryon asymmetry by ${\cal O}(M_{Pl}/m_{3/2})$.
This has been the major embarrassment of the hidden sector models
\cite{Coughlan:yk} or string theory \cite{deCarlos:1993jw}.  In our
case, on the other hand, they decay before BBN, 
a major improvement in supersymmetric cosmology.
  
There are remaining concerns, however.  For instance, if the moduli decay
copiously into superparticles, they will eventually cascade down to the
lightest supersymmetric particle (LSP) that would overclose the Universe 
(assuming the LSP is stable and has an electroweak-scale mass)
\cite{Moroi:1994rs,Kawasaki:1995cy}.\footnote{The branching fractions of
the moduli are certainly highly model-dependent and may not copiously
produce superparticles.}  One possibility is that the LSP has a mass 
below about 1 GeV and does not overclose the universe.  It was argued
\cite{Rajagopal:1990yx} that the axino, the fermionic superpartner of
the invisible axion, is a possible candidate of a light LSP in the
context of the gauge-mediated supersymmetry breaking.  In a typical
gravity-mediated supersymmetry breaking, the axino acquires a mass of
${\cal O}(m_{3/2})$ and this idea does not work \cite{Goto:1991gq}.  In
anomaly mediation, however, the axion lacks renormalizable couplings in
the infrared and hence its mass is indeed suppressed by the inverse
power of the decay constant.  The axino is therefore a viable candidate
for the LSP.  Yet another possibility is that the $R$-parity is broken
and the LSP decays.
  
Of course the moduli may simply not exist, or equivalently, may had
been fixed in a supersymmetric fashion.  Recent string theory
compactifications with anti-symmetric tensor field background together
with D-branes are shown to be capable of indeed doing so
\cite{Kachru:2002he,Gukov:2003cy}.  Hidden sector models can avoid
light moduli if the supersymmetry breaking is dynamical
\cite{Banks:1993en}, or of O'Raifeartaigh type instead of Polonyi-like
\cite{Joichi:1994ce,Greene:2002ku}.  There is a possibility that the
baryon asymmetry may be generated by the gravitino decay
\cite{Cline:1990bw}, which may be generalized to the moduli decay.

\subsection{Fine Tuning?}
\label{finetune-subsec}

The mass parameters in the Higgs sector appear alarmingly large because
of the large values of $\lambda$ and $f_t$, suggesting some fine-tuning among
the parameters is needed to obtain the small value of $v$.
Here we estimate the amount of fine-tuning for each case and find that
it is no worse than the MSSM with minimal supergravity.

The Higgs VEV $v$ is approximately solved by expanding
Eqs.~(\ref{constraint-1}) and (\ref{constraint-2}) in terms of $v^2/M^2$ as
follows:
\begin{eqnarray}
 v^2 &\sim&
\frac{m_1^2 + m_2^2}{2 \lambda^2}
\left[
1- \frac{m_1^2 m_2^2}{(A_\lambda S + \lambda m^2)^2}
\right] \nonumber \\
&\sim&
\frac{m_1^2 + m_2^2}{2 \lambda^2}
\left[
1- \left(
\frac{\lambda}{\lambda^2 + (4 \pi)^2 }
\right)^2
\frac{ m_1^2 m_2^2}{m^4}
\right] \ ,
\label{app-v}
\end{eqnarray}
where in the second expression we used an approximate solution of $S
\sim -C m^2 / m_S^2$ determined by Eq.~(\ref{constraint-3}).
Ignoring all the couplings except for $\lambda$, the prefactor of
$(m_1^2 + m_2^2)/(2 \lambda^2)$ is estimated to be $4 \lambda^2 M^2 +
S^2$ that is much larger than $v^2$ for the solutions discussed in
Sec.~\ref{potential-subsec}.  Therefore a cancellation between two terms 
in the parentheses is necessary.  It is important to remark that this
fine-tuning resulted because anomaly mediation is communicating
supersymmetry breaking to the soft masses.  Fine-tuning would be eliminated
if soft supersymmetry breaking were fixed independent of raising of 
$\lambda$, as originally found in \cite{fat}, but this is not possible 
in anomaly mediation because $m_1^2 + m_2^2$ contains a term proportional
to $\lambda^4$.

Fortunately the typical value of fine-tuning is ${\cal O}(1\%)$, no worse 
than the MSSM \cite{Casas:2003jx}.  This is quite surprising given 
that the spectrum 
is much heavier than the MSSM with gravity mediation.  The reason why the 
fine-tuning is smaller is because the Higgs VEV is mainly determined 
by the potential with coupling $\lambda$ and not the gauge couplings.
The corresponding formula in the MSSM is given by
\begin{eqnarray}
 v^2 \sim \frac{4 (m_1^2 + m_2^2) B^2 \mu^2}{(g_Y^2 + g_2^2)(m_1^2 - m_2^2)^2}
\left[
1- \frac{m_1^2 m_2^2}{B^2 \mu^2}
\right]\ .
\end{eqnarray}
Clearly a more accurate cancellation is necessary for the MSSM as 
compared to Eq.~(\ref{app-v}) because of the small gauge coupling 
constants, even when all the supersymmetry breaking parameters are 
of the same order.


\section{Conclusions}
\label{conclusions-sec}

We found a stable electroweak symmetry breaking vacuum with
supersymmetry breaking communicated via UV insensitive anomaly
mediation.  In the MSSM we found that this occurs when $\tan\beta \sim
0.3$; no solutions for $\tan\beta > 1$ were found.  This implies the top
Yukawa coupling blows up slightly above the supersymmetry breaking scale.  The
MSSM with UV insensitive anomaly mediation is therefore UV incomplete,
even though (by definition) this does not affect the predictions on the
supersymmetry breaking parameters.  We searched the parameter space of
the NMSSM and were unable to find a viable set of parameters due to the
prediction of tiny chargino masses that are ruled out by direct search
experiments.

We did find a stable vacuum and viable spectrum for a model with a
modified (NMSSM-like) superpotential in which the the cubic singlet term
is replaced with a term linear in the singlet.  Two distinct regions of
parameter space were found: $\tan\beta \lsim 1$ and $\tan\beta \gtrsim
5$.  The parameter region with $\tan\beta \lsim 1$ requires the $S H_1
H_2$ coupling to be $\lambda \gtrsim 1.5$, that is remarkably just
what is expected in the ``Fat Higgs'' model that generates this
NMSSM-like superpotential dynamically.  The first UV complete,
UV insensitive supersymmetry breaking model with a stable electroweak
symmetry breaking vacuum is therefore the Fat Higgs model with
UV insensitive anomaly mediation.  

We also found a viable parameter region with $\tan \beta \gtrsim 5$.
In this case, $\lambda \simeq 1.0$ and the Higgs is ``less fat.''
Since $\tan \beta$ is moderately large, the Higgs potential is similar
to that in the MSSM, and hence the large radiative corrections to the
lightest Higgs mass are necessary to satisfy the LEP bound.  The large
$\tan \beta$ implies small Higgsino masses and the lightest neutralino
is Higgsino-rich.  It tends to come around $m_Z/2$ and an additional
contribution to the invisible decay width of the $Z$-boson can be a
problem.  Nonetheless there are viable parameter sets.

There are numerous implications for collider phenomenology and
cosmology.  In the region $\tan\beta \lsim 1$, the LHC is expected to
discover supersymmetry through missing energy and dilepton signatures
of squarks and gluinos.  The trilepton signatures of
charginos/neutralinos are viable at Tevatron and LHC unlike most
anomaly-mediated models in the literature that make use of UV
sensitive universal scalar mass.  The lightest (SM-like) Higgs will
decay mostly into $WW$ and $ZZ$ making it easy to find at the LHC.
Finally, the abundance of the lightest supersymmetric particle, a
neutralino that is mostly a singlino, could well be in the right range
if its mass is of order $m_Z/2$.  The combination of LHC and LC would
allow for a quantitative test of the predicted gaugino mass relation.
In the other region $\tan\beta \gtrsim 5$, the squarks and gluinos may
be beyond the reach of LHC.  On the other hand, both wino-like and
Higgsino-like charginos and neutralinos are light and can be looked
for.

The model predicts that the gravitino is heavy, $\sim 100$~TeV,
solving its cosmological problem and allowing thermal leptogenesis to
be a realistic possibility.

Finally, note that we have not attempted a full parameter scan of the model.  
We believe it would be useful to perform such a scan, combined with more 
detailed analysis of the precision electroweak constraint, the neutralino 
abundance, and the collider signatures.

\section{Acknowledgments}

RK is the Marvin L. Goldberger Member and GDK is a Frank and Peggy 
Taplin Member; both of us thank these individuals for their generous 
support of the School of Natural Sciences. 
This work was supported by the Institute for Advanced Study,
funds for Natural Sciences, as well as in part by the DOE under
contracts DE-FG02-90ER40542 and DE-AC03-76SF00098 and in part by NSF
grant PHY-0098840.


\end{document}